\documentstyle[psfig,12pt,axodraw]{article}
\begin{document}
\baselineskip 18pt
\def\today{\ifcase\month\or
 January\or February\or March\or April\or May\or June\or
 July\or August\or September\or October\or November\or December\fi
 \space\number\day, \number\year}
\def\thebibliography#1{\section*{References\markboth
 {References}{References}}\list
 {[\arabic{enumi}]}{\settowidth\labelwidth{[#1]}
 \leftmargin\labelwidth
 \advance\leftmargin\labelsep
 \usecounter{enumi}}
 \def\newblock{\hskip .11em plus .33em minus .07em}
 \sloppy
 \sfcode`\.=1000\relax}
\let\endthebibliography=\endlist
\def\gsim{{\mathop >\limits_\sim}}
\def\lsim{{\mathop <\limits_\sim}}
\def\PL{{1-\gamma_5\over 2}}
\def\PR{{1+\gamma_5\over 2}}
\def\r2{\sqrt 2}
\def\a#1{\alpha_#1}
\def\sw2{\sin^2\theta_W}
\def\tw{\tan\theta_W}
\def\v#1{v_#1}
\def\tb{\tan\beta}
\def\c2b{\cos 2\beta}
\def\w{\omega}
\def\x{\chi}
\def\g{\tilde g}
\def\st{\tilde t}
\def\sb{\tilde b}
\def\HC{H^\pm}
\def\m#1{{\tilde m}_#1}
\def\mg{{\tilde m}_g}
\def\mH{m_H} 
\def\Mh{M_{H^\pm}}
\def\mgr{m_{3/2}}
\def\mw{\tilde m_\omega}
\def\mx{\tilde m_\chi}
\def\M{\tilde M}
\def\Mst{\tilde M_t}
\def\Msb{\tilde M_b}
\def\laba{\log \Biggl| \frac{S+T}{S-T}\Biggr|}
\def\labb{\log \Biggl| \frac{S'+T}{S'-T}\Biggr|}
\begin{titlepage}
\hspace*{10.0cm}ICRR-Report-428-98-24 
   
\hspace*{10.0cm}OCHA-PP-119
\  \
\vskip 0.5 true cm 
\begin{center}
{\large {\bf Decay Rate Asymmetry of Top Squark}}  
\vskip 2.0 true cm
\renewcommand{\thefootnote}
{\fnsymbol{footnote}}
Mayumi Aoki $^1$ \footnote{Research Fellow of the Japan Society 
for the Promotion of Science.} and Noriyuki Oshimo $^2$ \\
\vskip 0.5 true cm 
{$^1$ \it Graduate School of Humanities and Sciences,  
Ochanomizu University}  \\
{\it Otsuka 2-1-1, Bunkyo-ku, Tokyo 112-8610, Japan}  \\
{$^2$ \it Institute for Cosmic Ray Research, 
University of Tokyo} \\
{\it Midori-cho 3-2-1, Tanashi, Tokyo 188-8502, Japan}  \\
\end{center}

\vskip 4.0 true cm

\centerline{\bf Abstract}
\medskip
     We discuss a decay rate asymmetry of the top squark, 
which is induced by a new source of $CP$ violation intrinsic  
in the supersymmetric standard model.
Although new sources of $CP$ violation in this model are severely 
constrained from the electric dipole moment of the neutron, 
an unsuppressed $CP$-violating phase can still coexist with 
a top squark whose mass is accessible by near-future colliders.  
Then the dominant decay mode of the top squark has 
a width different from its $CP$ conjugate process.  
The magnitude of this $CP$ asymmetry 
becomes of order $10^{-3}$. 

\vskip 1.0 true cm

\noindent
{\it PACS}:  11.30.Er, 12.60.Jv, 14.80.Ly  \\
{\it Keywords}:  $CP$ violation; Supersymmetry; Top squark

\end{titlepage}

     In the standard model (SM) a physical complex phase in 
quark mass matrices is, aside from the strong $CP$ phase,  
the only source of $CP$ violation.  
This phase can explain $CP$ violation in the $K^0$-$\bar K^0$ system,  
whereas other sizable $CP$-violating phenomena are not 
expected except in the $B^0$-$\bar B^0$ system \cite{cprev}.  
On the other hand, various models 
beyond the SM contain new sources of $CP$ violation.   
These effects may be observed in reactions where the 
SM does not predict $CP$ violation.  For instance, near-future 
collider experiments, such as LHC and NLC, enable 
precise measurements of certain reactions, which could 
provide an opportunity of detecting new $CP$-violating phenomena.  

     The supersymmetric standard model (SSM) contains 
a new source of $CP$ violation in the top-squark system \cite{topssm}.   
We have recently studied their effects on 
the decay of the top quark \cite{topdecay}.  
Taking into account the constraints from the electric dipole 
moment (EDM) of the neutron, 
it was shown that a rate asymmetry between the decays 
$t\rightarrow bW^+$ and $\bar t\rightarrow \bar b W^-$ 
can be induced at a magnitude of order $10^{-3}$.  
However, this asymmetry is generated only if there 
exists a top squark lighter than the top quark.  
If this condition is not satisfied, 
other phenomena have to be invoked to search 
for $CP$ violation arising from the top-squark system.  
Since $CP$ violation by the top-squark interactions 
may be able to explain baryon asymmetry of the universe \cite{baryon}, 
it would be of interest to examine this mechanism of $CP$ violation.  
 
     In this letter, we discuss $CP$ violation which is induced by 
the top squark heavier than the top quark.  If a top squark is sufficiently 
heavy, its dominant decay modes are tree-level two-body decays  
$\st\rightarrow b\omega$ and $\st\rightarrow t\x$, 
where $\w$ and $\x$ denote a chargino and a neutralino, respectively.  
These final states are also produced at the one-loop level by  
final state interactions.   
If the interactions of the top squark violate $CP$ invariance, 
these decays are expected to have rates different from their $CP$ conjugate 
processes, which are measured by the asymmetries  
\begin{eqnarray}
A_{CP}^b &=& \frac{\Gamma(\st\rightarrow b\w^+)
                                 -\Gamma(\st^*\rightarrow \bar b\w^-)}
       {\Gamma(\st\rightarrow b\w^+)+\Gamma(\st^*\rightarrow \bar b\w^-)},  
\label{basy} \\  
A_{CP}^t &=& \frac{\Gamma(\st\rightarrow t\x)
                                 -\Gamma(\st^*\rightarrow \bar t\x)}
       {\Gamma(\st\rightarrow t\x)+\Gamma(\st^*\rightarrow \bar t\x)}.    
\label{tasy}
\end{eqnarray} 
Owing to $CPT$ invariance, the decay widths satisfy the relation 
\begin{equation}
\Gamma(\st\rightarrow b\w^+)-\Gamma(\st^*\rightarrow \bar b\w^-)=
-\left\{\Gamma(\st\rightarrow t\x)-\Gamma(\st^*\rightarrow \bar t\x)\right\},  
\label{cpt}
\end{equation}  
making the total width of the top squark to be the same as that 
of the anti-top squark.  
We calculate the asymmetry $A_{CP}^b$.  
Its magnitude can become of order $10^{-3}$, which may be accessible at  
near-future colliders.    
 
     We first briefly review the model \cite{ssmrev}.   
The SSM contains two new $CP$-violating phases.   
These phases can be taken, without loss of generality, for those of 
the dimensionless coupling constant 
$A$ in the trilinear terms of the scalar fields and  
the Higgsino mass parameter $\mH$ in the bilinear term 
of the Higgs superfields, 
\begin{equation}
    A = |A|\exp(i\alpha),  \quad   \mH = |\mH|\exp(i\theta). 
\label{cpphase}
\end{equation} 
The mass-squared matrix $M^2_t$ for the top squarks have 
these phases, which is given by
\begin{equation}
    \lefteqn{M^2_t =} \hspace{9cm}
\label{stopmass} 
\end{equation}
\[
 \left(\matrix{m_t^2+\c2b(\frac{1}{2}-\frac{2}{3}\sw2)M_Z^2+\M_{tL}^2 &
                            m_t (\cot\beta\mH + A^*\mgr) \cr
         m_t (\cot\beta\mH^* + A\mgr) &
                   m_t^2 + \frac{2}{3}\c2b\sw2 M_Z^2+\M_{tR}^2}
           \right),
\]
with $\mgr$ and $\tb$ being respectively the gravitino mass 
and the ratio of the vacuum expectation values of the Higgs bosons.  
The mass-squared parameters $\M_{tL}^2$ and $\M_{tR}^2$ 
are expressed as 
\begin{equation}
 \M_{tL}^2  = \M_q^2-cm_t^2,  \quad \M_{tR}^2 = \M_q^2-2cm_t^2,  
\end{equation}
where $\M_q$ is of order $\mgr$ and  
represents approximately squark masses for the first two generations. 
The dimensionless constant $c$ stands for an  
amount of quantum corrections through   
the Yukawa interactions proportional to the top-quark mass, 
with $c = 0.1-1$.  

     The mass-squared matrix $M^2_t$ could become an origin of sizable 
$CP$ violation, if the imaginary parts of the off-diagonal elements  
are comparable with the real parts and the diagonal elements.   
This condition requires that $\M_q$ be not much larger than $m_t$,   
and at least either $\theta$ or $\alpha$ be of order unity.   
However, these parameters are severely constrained by the EDM of the neutron.  
An unsuppressed magnitude for $\theta$ and a value of $\M_q$  
smaller than 1 TeV give a too large magnitude to the neutron EDM 
through one-loop diagrams mediated by charginos and squarks \cite{edm}.   
On the other hand, the experimental upper bound on this EDM 
is compatible with $\alpha\sim 1$ and $\M_q < 1$ TeV, provided 
that the gluinos are sufficiently heavy \cite{topdecay}.  
We thus assume that $\theta$ is much smaller than unity  
and $\alpha$ alone has a nonnegligible value.    

\begin{figure}
\caption{The one-loop diagrams for the decay of a top squark 
into a bottom quark and a chargino.     
  }
\label{oneloop}

\vspace{2cm}

%
\begin{center}\begin{picture}(400,100)(0,0)
\DashLine(23,100)(70,100){5}
\Line(70,100)(130,130)
\Line(70,100)(130,70)
\DashLine(130,130)(130,70){5}
\Line(130,130)(180,130)
\Line(130,70)(180,70)
\Text(17,100)[]{$\st$}
\Text(100,125)[]{$\x$}
\Text(100,75)[]{$t$}
\Text(150,100)[]{$\HC$}
\Text(188,130)[]{$\w$}
\Text(188,70)[]{$b$}
\Text(105,50)[]{(a)}
\DashLine(223,100)(270,100){5}
\Line(270,100)(330,130)
\Line(270,100)(330,70)
\Photon(330,130)(330,70){5}{4}
\Line(330,130)(380,130)
\Line(330,70)(380,70)
\Text(217,100)[]{$\st$}
\Text(300,125)[]{$\x$}
\Text(300,75)[]{$t$}
\Text(350,100)[]{$W$}
\Text(388,130)[]{$\w$}
\Text(388,70)[]{$b$}
\Text(305,50)[]{(b)}
\DashLine(23,0)(70,0){5}
\Line(70,0)(130,30)
\Line(70,0)(130,-30)
\DashLine(130,30)(130,-30){5}
\Line(130,30)(180,30)
\Line(130,-30)(180,-30)
\Text(17,00)[]{$\st$}
\Text(100,25)[]{$t$}
\Text(100,-25)[]{$\x$}
\Text(150,0)[]{$\sb$}
\Text(188,30)[]{$\w$}
\Text(188,-30)[]{$b$}
\Text(105,-50)[]{(c)}
\end{picture}
\end{center}

\vspace{4cm}
\end{figure}
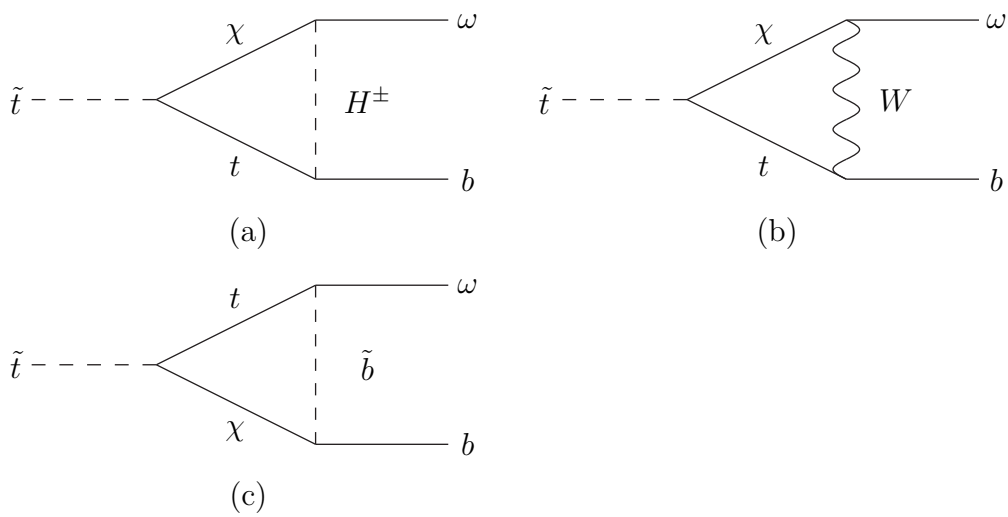

     A nonvanishing value for the asymmetry $A_{CP}^b$
in Eq. (\ref{basy}) is generated, if the decay
$\st\rightarrow t\x$, in addition to the decay $\st\rightarrow b\w$, 
is allowed kinematically.  
The produced top quark and neutralino can become a
bottom quark and a chargino by exchanging charged Higgs 
bosons $\HC$, $W$ bosons, and bottom squarks $\sb$, as shown 
in Fig. \ref{oneloop}.
The interferences of these one-loop diagrams with the tree diagram
make the rate of the decay
$\st\rightarrow b\w^+$ different from that of the decay 
$\st^*\rightarrow\bar b\w^-$.
The relevant interaction Lagrangian for $\st$, $\sb$, and $W$ is given by 
\begin{eqnarray}
 \cal L &=&  i\frac{g}{\r2}\sb_k^\dagger \bar{\w_i^c} 
              \left(A_{Li}^k\PL+A_{Ri}^k\PR\right) t  
       + i\frac{g}{\r2}\st_k^\dagger\bar{\w_i}
              \left(B_{Li}^k\PL+B_{Ri}^k\PR\right) b  \nonumber    \\
 &+& i\frac{g}{\r2}\st_k^\dagger\bar{\x_j}
              \left(F_{Lj}^k\PL+F_{Rj}^k\PR\right) t     
       + i\frac{g}{\r2}\sb_k^\dagger\bar{\x_j}
              \left(G_{Lj}^k\PL+G_{Rj}^k\PR\right) b \nonumber \\ 
 &+& \frac{g}{\r2}W_\mu^\dagger\bar{\x_j}\gamma^\mu
               \left(H_{Lji}\PL+H_{Rji}\PR\right)\w_i   +{\rm H.c.}.   
\label{lag1}
\end{eqnarray}
For the definitions of various coefficients, we refer to Ref. \cite{topdecay}.  
The interaction Lagrangian for $\HC$ is given by 
\begin{eqnarray}
\cal L &=&  \frac{g}{\r2}H^+\bar{\x_j}
                  \left(I_{Lji}\PL+I_{Rji}\PR\right) \w_i     \nonumber \\
      &+& \frac{g}{\r2}H^+\bar{t}
                  \left(K_{L}\PL+K_{R}\PR\right)b  +{\rm H.c.},  
\label{lag2} 
\end{eqnarray}
where coefficients are defined as 
\begin{eqnarray}
   I_{Lji} &=& -\sin\beta\{(\tw N_{1j}+N_{2j})C_{L2i}-\r2 N_{3j}C_{L1i}\}, 
                                \nonumber \\ 
 I_{Rji}& =& \cos\beta\{(\tw N_{1j}^*+N_{2j}^*)C_{R2i}+\r2 N_{4j}^*C_{R1i}\},  
                            \nonumber \\ 
  K_L &=& \cot\beta\frac{m_t}{M_W}, \quad  K_R=\tb\frac{m_b}{M_W},   
\end{eqnarray}   
$C_L$, $C_R$, and $N$ being the unitary matrices which 
diagonalize the chargino and neutralino mass matrices \cite{topdecay}.   
The SSM parameters appearing in our analyses 
are $\tb$, $A$, $\mH$, $\M_q$, $\mgr$, $c$, the SU(2) 
gaugino mass $\m2$, and the charged Higgs-boson mass $\Mh$.
Although these parameters are not all independent of each other,
they can have various sets of values
depending on assumptions for underlying models.
For simplicity, we take those parameters independent and
assume only rough constraints coming from theoretical and
experimental considerations.

     We now consider the decay of the lighter top squark into 
the bottom quark and the lighter chargino $\st_1\rightarrow b\w_1^+$ 
and its $CP$ conjugate process.  
The decay rate asymmetry for these processes are obtained as
\begin{eqnarray}
    A_{CP}^b &=&     
  \frac{\alpha_2}{2}\biggl[\biggl\{(|B_{L1}^1|^2+|B_{R1}^1|^2)
 (\M_{t1}^2-m_b^2-\tilde m_{\w 1}^2)  \nonumber \\
           &-& 4{\rm Re}(B_{L1}^1B_{R1}^{1*})m_b\tilde m_{\w 1}\biggl\}
    \sqrt{\lambda(\M_t^2,\tilde m_{\w 1}^2,m_b^2)}\biggl]^{-1} 
     \left(T^a + T^b +T^c \right),   
\label{asy}
\end{eqnarray}
where the kinematic function $\lambda(a,b,c)$ is defined by 
\begin{equation}
\lambda(a,b,c)=a^2+b^2+c^2-2ab-2bc-2ca.
\end{equation}
The contributions of the diagrams (a), (b), and (c) in Fig. \ref{oneloop}   
are represented by $T^a$, $T^b$, and $T^c$, respectively.  
The terms $T^a$ and $T^b$ are written as  
\begin{eqnarray}
T^a &=& \sum_j\sum_{n=1}^8{\rm Im}(X_{1j1}^n)
      I_n(\tilde m_{\w 1},\tilde m_{\x j},\M_{t1}),   
\label{ta}        \\
T^b &=& \sum_j\sum_{n=1}^8{\rm Im}(Y_{1j1}^n)
      J_n(\tilde m_{\w 1},\tilde m_{\x j},\M_{t1}),   
\label{tb}
\end{eqnarray}
where coefficients $X_{1j1}^n$, $Y_{1j1}^n$ and functions
$I_n(\tilde m_{\w 1},\tilde m_{\x j},\M_{t1})$,
$J_n(\tilde m_{\w 1},\tilde m_{\x j},\M_{t1})$
are given in Appendix.
The term $T^c$ is obtained from $T^a$ by appropriately changing the 
coefficients and the charged Higgs-boson mass $\Mh$.  
The sum for the intermediate neutralinos should be done for 
those which satisfy the kinematical 
condition $\M_{t1} > m_t + \tilde m_{\x j}$.   

     In Fig. 2 the absolute value of $A_{CP}^b$ is shown 
as a function of the SU(2) gaugino mass $\m2$ 
for $\M_q=400$ GeV and $\alpha=\pi/4$.
Four curves correspond to four sets of parameter values for $c$ and $\tb$ 
listed in Table\ \protect\ref{ttasy},  
where the lighter top-squark mass is also given.   
The other parameters are taken as $|A|\mgr=\M_q$, $|\mH|=100$ GeV, $\theta=0$,  
and $\Mh=100$ GeV. The parameter $\m2$, together with 
$\mH$ and $\tb$, determines the masses of the charginos and neutralinos.   
In the mass ranges where curves are not drawn, the lighter chargino mass 
becomes smaller than 66 GeV, which is ruled out 
by LEP2 experiments \cite{pdg}.  
In Table\ \protect\ref{cnmass} the masses of the lighter chargino 
and the lighter two neutralinos are shown for $\m2=500$ GeV.  
Assuming grand unified models, the gluino mass is also determined by $\m2$. 
For $\m2\gsim 500$ GeV, the gluinos are heavier than 2 TeV and  
the value of the neutron EDM lies within its experimental bound. 

     The asymmetry $A_{CP}^b$ has a magnitude of order $10^{-3}$
in a region of parameter space where $CP$ violation becomes 
maximal at the Lagrangian level.   
For a larger value of $|\mH|$ or $\M_q$, $|A_{CP}^b|$ is smaller.   
The mass of the charged Higgs boson does not affect much the asymmetry.  
In most region of parameter space in Fig. 2, $\st_1$  
dominantly decays into $b\w^+_1$, $t\x_1$, and $t\x_2$.  
The branching ratios of these decays at $\m2=500$ GeV are shown in 
Table\ \protect\ref{branching}.  
For the detection of an asymmetry $A_{CP}^b\sim 10^{-3}$, 
a necessary number for pairs of $\st\st^*$ is  
of order $10^{6}$.
At LHC, the $\st\st^*$ pairs are expected to be produced at a rate of
order $10^5$.  If its luminosity is upgraded by one order of 
magnitude, the asymmetry will be within the  reach of detectability.   

     The interactions which induce the rate asymmetry
between the decays $\st \rightarrow b\w^+$ and 
$\st^*\rightarrow \bar b\w^-$
also yield a rate asymmetry between the decays
$\st \rightarrow t\x$ and $\st^* \rightarrow \bar t\x$, satisfying 
the relation in Eq. (\ref{cpt}).  
As seen in Table\ \protect\ref{branching}, the width of 
$\st_1\rightarrow t\x_1$ or $\st_1\rightarrow t\x_2$ is generally  
several times smaller than that of $\st_1\rightarrow b\w^+_1$, 
and accordingly the former decay rate asymmetry becomes larger than the latter 
by the same order of magnitude.   
It will also be possible to examine $CP$ violation 
through the decay into $t\x_1$ or $t\x_2$.   
However, the relation in Eq. (\ref{cpt}) makes the detection of a decay rate
asymmetry a little involved.  
The top quark decays into $b$ and $W$,  while the chargino 
decays through $\x_1$ and $W^*$, 
where $W^*$ denotes the virtual state of the $W$ boson.  
Consequently, the top-squark decays $\st_1\rightarrow b\w^+_1$ and 
$\st_1\rightarrow t\x_1$ have the same particles in their final states.  
In order to measure the asymmetry, these decays have 
to be distinguished.  
Such a distinction can be made by e.g. examining the energy of the $b$ quark.  
On the other hand, the decay $\st_1\rightarrow t\x_2$ shows a different 
topology, because of the subsequent decay of the second 
lightest neutralino $\x_2$.   

     In summary, we have studied the decay rate asymmetry of the  
lighter top squark.  
By this asymmetry, $CP$ violation intrinsic in the SSM can be probed.  
If the gluinos have a mass larger than 1 TeV, 
a new $CP$-violating phase contained in the top-squark mass-squared matrix
is not constrained much from the EDM of the neutron.  
For a top-squark mass around 300 GeV with the $CP$-violating phase 
of order unity, the asymmetry is of 
order $10^{-3}$, which may be detectable in the near future. 

\vskip 1.0 true cm 

     The work of M.A. is supported in part by the Grant-in-Aid for Scientific 
Research from the Ministry of Education, Science and Culture, Japan.  

\section*{Appendix}

     The coefficients $X_{ijk}^n$ and the functions  
$I_n(\tilde m_\w,\tilde m_\x,\M_t)$ in Eq. (\ref{ta}) are defined by 
\begin{eqnarray} 
  X_{ijk}^1 &=& K_{L}^*B_{Li}^{k}F_{Lj}^{k*}I_{Lji}  
             + K_{R}^*B_{Ri}^{k}F_{Rj}^{k*}I_{Rji},  \nonumber \\ 
  X_{ijk}^2 &=& K_{L}^*B_{Li}^{k}F_{Rj}^{k*}I_{Lji}  
             + K_{R}^*B_{Ri}^{k}F_{Lj}^{k*}I_{Rji},  \nonumber \\ 
  X_{ijk}^3 &=& K_{L}^*B_{Ri}^{k}F_{Lj}^{k*}I_{Lji}  
             + K_{R}^*B_{Li}^{k}F_{Rj}^{k*}I_{Rji},  \nonumber \\ 
  X_{ijk}^4 &=& K_{L}^*B_{Ri}^{k}F_{Rj}^{k*}I_{Lji}  
             + K_{R}^*B_{Li}^{k}F_{Lj}^{k*}I_{Rji},  \nonumber \\ 
  X_{ijk}^5 &=& K_{R}^*B_{Li}^{k}F_{Lj}^{k*}I_{Lji}  
             + K_{L}^*B_{Ri}^{k}F_{Rj}^{k*}I_{Rji},  \nonumber \\ 
  X_{ijk}^6 &=& K_{R}^*B_{Li}^{k}F_{Rj}^{k*}I_{Lji}  
             + K_{L}^*B_{Ri}^{k}F_{Lj}^{k*}I_{Rji},  \nonumber \\ 
  X_{ijk}^7 &=& K_{R}^*B_{Ri}^{k}F_{Lj}^{k*}I_{Lji}  
             + K_{L}^*B_{Li}^{k}F_{Rj}^{k*}I_{Rji},  \nonumber \\ 
  X_{ijk}^8 &=& K_{R}^*B_{Ri}^{k}F_{Rj}^{k*}I_{Lji}  
             + K_{L}^*B_{Li}^{k}F_{Lj}^{k*}I_{Rji},  \nonumber  
\end{eqnarray}
\begin{eqnarray}
 I_1 &=& \frac{1}{2}\mw m_t
\left\{T+(m_t^2+\mw^2-\Mst^2-\Mh^2)\laba\right\}, \nonumber \\
 I_2 &=& \frac{1}{2}\mw\mx 
\left\{T+(m_t^2+m_b^2-\Mh^2)\laba\right\}, \nonumber \\
 I_3 &=& \frac{1}{2}m_b m_t
\left\{T+(\mx^2+\mw^2-\Mh^2)\laba\right\}, \nonumber \\
 I_4 &=& \frac{1}{2}m_b\mx 
\left\{T+(\mx^2+m_b^2-\Mst^2-\Mh^2)\laba\right\}, \nonumber \\
 I_5 &=& -\frac{1}{2}\mw m_b(\Mst^2-m_t^2-\mx^2)\laba,  \nonumber \\
 I_6 &=& m_b m_t \mw\mx\laba,     \nonumber  \\
 I_7 &=& \frac{1}{2} 
\left\{\Mst^2T+(-\Mst^2\Mh^2+\mw^2 m_t^2+m_b^2\mx^2)\laba\right\}, \nonumber \\
 I_8 &=& -\frac{1}{2}\mx m_t(\Mst^2-m_b^2-\mw^2)\laba,  \nonumber 
\end{eqnarray}
where
\begin{eqnarray}
S &=& \frac{1}{\Mst^2}(\Mst^2 +\mw^2 -m_b^2 )(\Mst^2+\mx^2 -m_t^2)
        -2(\mx^2+\mw^2-\Mh^2), \nonumber \\
T &=& \frac{1}{\Mst^2}\sqrt{\lambda(m_b^2,\Mst^2,\mw^2) 
                                     \lambda(m_t^2,\Mst^2,\mx^2)}.  
\nonumber 
\end{eqnarray}

     The coefficients $Y_{ijk}^n$ and the functions  
$J_n(\tilde m_\w,\tilde m_\x,\M_t)$ in Eq. (\ref{tb}) are defined by 
\begin{eqnarray} 
 & & Y_{ijk}^1 = B_{Li}^{k}F_{Lj}^{k*}H_{Lji},  \quad  
     Y_{ijk}^2 = B_{Li}^{k}F_{Lj}^{k*}H_{Rji},  \quad  
     Y_{ijk}^3 = B_{Li}^{k}F_{Rj}^{k*}H_{Lji},  \nonumber \\ 
 & & Y_{ijk}^4 = B_{Li}^{k}F_{Rj}^{k*}H_{Rji},  \quad  
     Y_{ijk}^5 = B_{Ri}^{k}F_{Lj}^{k*}H_{Lji},  \quad  
     Y_{ijk}^6 = B_{Ri}^{k}F_{Lj}^{k*}H_{Rji},  \nonumber \\ 
 & & Y_{ijk}^7 = B_{Ri}^{k}F_{Rj}^{k*}H_{Lji},  \quad  
     Y_{ijk}^8 = B_{Ri}^{k}F_{Rj}^{k*}H_{Rji},  \nonumber  
\end{eqnarray}
\begin{eqnarray}
 J_1 &=& \frac{1}{2}\mw\mx 
   \biggl[\frac{2M_W^2+m_t^2+m_b^2}{M_W^2}T \nonumber \\
     & & +\left\{m_b^2+m_t^2-2M_W^2
  +\frac{(m_b^2-m_t^2)^2}{M_W^2}\right\}\labb\biggr],  \nonumber \\
 J_2 &=& -\frac{m_b^2\mx^2+m_t^2\mw^2}{2M_W^2}T 
     -\frac{1}{2}\biggl[2(\Mst^2-\mx^2-m_t^2)(\Mst^2-\mw^2-m_b^2) \nonumber \\
  & &    -m_b^2\mx^2-m_t^2\mw^2 +\frac{1}{M_W^2}
     \biggl\{(m_b^2\mx^2-m_t^2\mw^2)(\mx^2-\mw^2+m_b^2-m_t^2) \nonumber \\
   & &   +(m_b^2-m_t^2)(\mw^2-\mx^2)\Mst^2\biggr\}\biggr]\labb,  \nonumber \\
 J_3 &=& \frac{1}{2}\mw m_t\biggl[\frac{2M_W^2-\Mst^2+\mx^2+m_b^2}{M_W^2}T 
               +\biggl\{2(\mw^2-M_W^2+m_t^2) \nonumber \\
     & &   -\Mst^2-m_b^2-\mx^2 
 +\frac{(m_b^2-m_t^2)(\mw^2-\mx^2)}{M_W^2}\biggr\}\labb\biggr], \nonumber \\ 
 J_4 &=& \frac{\mx m_t(\Mst^2-m_b^2-\mw^2)}{2M_W^2}
                  \left(T+3M_W^2\labb \right),  \nonumber \\  
 J_5 &=& \frac{1}{2}\mx m_b
   \biggl[\frac{2M_W^2-\Mst^2+\mw^2+m_t^2}{M_W^2}T 
       +\biggl\{2(\mx^2-M_W^2+m_b^2) \nonumber \\
    & & -\Mst^2-m_t^2-\mw^2
  +\frac{(m_b^2-m_t^2)(\mw^2-\mx^2)}{M_W^2}\biggr\}\labb\biggr],  \nonumber \\  
 J_6 &=& \frac{\mw m_b(\Mst^2-m_t^2-\mx^2)}{2M_W^2}
                  \left(T+3M_W^2\labb \right),  \nonumber \\
 J_7 &=& \frac{1}{2}m_t m_b 
   \biggl[\frac{2M_W^2+\mx^2+\mw^2}{M_W^2}T  \nonumber \\
     & & +\biggl\{\mx^2+\mw^2-2M_W^2
  +\frac{(\mx^2-\mw^2)^2}{M_W^2}\biggr\}\labb\biggr],  \nonumber \\
 J_8 &=& -\frac{m_b m_t \mw\mx}{M_W^2}\left(T+3M_W^2\labb \right),  
        \nonumber 
\end{eqnarray}
where $S'$ is derived from $S$ by changing $\Mh$ to $M_W$.  

\newpage

\newpage 

\begin{table}
\caption{The parameter values for curves 
            (i)--(iv) in Fig.\ \protect\ref{tasy2}. 
        The lighter top-squark mass is also given.}
\label{ttasy}

\vspace{1cm}
\begin{center}
\begin{tabular}{cccc}
    & $c$  & $\tb$ & $\M_{t1}$ (GeV) \\
\hline  
 (i)     & 0.1 &  2 &  328   \\
 (ii)    & 0.1 & 10 &  335   \\
 (iii)   & 0.5 &  2 &  296   \\  
 (iv)    & 0.5 & 10 &  304   \\
\end{tabular} 
\end{center}
\end{table}

\begin{table}
\caption{The chargino and neutralino masses at $\m2=500$ GeV 
            for the parameter values of curves (i)--(iv) 
             in Fig.\ \protect\ref{tasy2}.}
\label{cnmass}

\vspace{1cm}
\begin{center}
\begin{tabular}{ccccc}
    & $\tilde m_{\w 1}$  & $\tilde m_{\x 1}$ & $\tilde m_{\x 2}$ & (GeV) \\
\hline  
 (i),(iii)     & 87 & 77 & 102  & \\
 (ii),(iv)     & 95 & 85 & 107  & \\
\end{tabular} 
\end{center}
\end{table}

\begin{table}
\caption{The branching ratios 
            for the four sets of parameter values  
            in Table\ \protect\ref{ttasy} with $\m2=500$ GeV.}
\label{branching}

\vspace{1cm}
\begin{center}
\begin{tabular}{cccc}
    & $\st_1\rightarrow b\w^+_1$  & $\st_1\rightarrow t\x_1$ & 
                                        $\st_1\rightarrow t\x_2$  \\
\hline  
 (i)     & 0.59 &  0.14 &  0.27   \\
 (ii)    & 0.59 & 0.13 &  0.28  \\
 (iii)  & 0.72 &   0.09 &  0.19   \\  
 (iv)   & 0.71 &   0.09 &  0.20   \\
\end{tabular} 
\end{center}
\end{table}

\newpage   

%
\pagebreak

%
\begin{figure}
\caption{The decay rate asymmetry  
   as a function of $\m2$.  
   Four curves correspond to four sets of values for $c$ 
   and $\tb$ in Table\ \protect\ref{ttasy}.  
   }
\label{tasy2}

\vspace{1.5cm}
\psfig{file=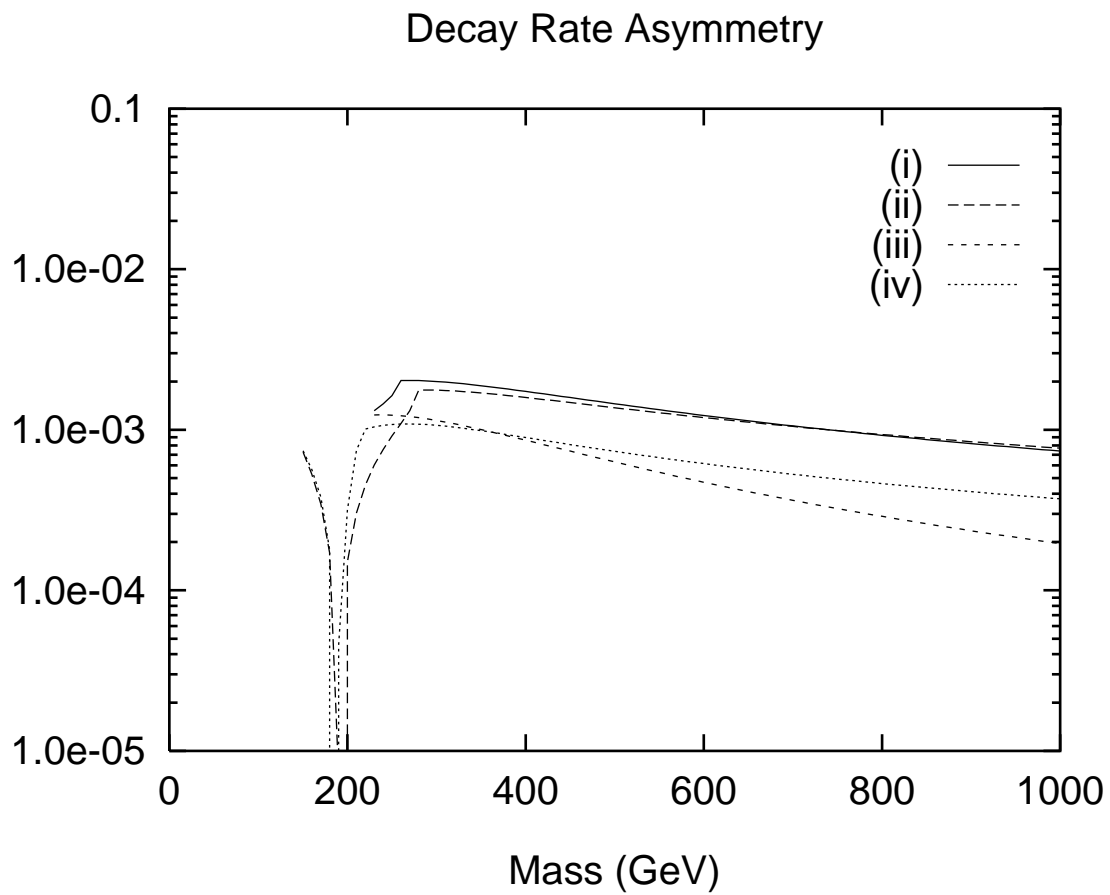}
\end{figure}
\vspace{2cm}

\end{document}